\documentclass[11pt]{article}
\RequirePackage{rotating}
\usepackage{subcaption}
\usepackage{float}
\usepackage[affil-it]{authblk}
\usepackage{booktabs}
\usepackage{amssymb}

\title{$\widetilde{\tau}$ searches at the ILC}
\author[1]{M.T. N{\'u}{\~n}ez Pardo de Vera\thanks{Corresponding author: maria-teresa.nunez-pardo-de-vera@desy,de}}
\author[1]{M. Berggren}
\author[1]{J. List}

\affil[1]{%
  DESY}

\date{May 2021}

\begin{document}

\maketitle

\begin{abstract}
The direct pair-production of the superpartner of the $\tau$-lepton, the $\widetilde{\tau}$,  is one
of the most interesting channels to search for SUSY in. First of all, the $\widetilde{\tau}$ is
likely to be the lightest of the scalar leptons. Secondly the
signature of $\widetilde{\tau}$ pair production signal events is one of the experimentally most difficult
ones, thereby constituting the ``worst'' possible scenario for SUSY searches.
The current model-independent $\widetilde{\tau}$ limits comes from analyses performed at
LEP but they suffer from the limited energy of this facility. Limits obtained at the LHC do
extend to higher masses, but they are only valid under strong assumptions.
ILC, a future electron-positron collider with
energy up to 500 GeV and upgrade capability\footnote{The initial ILC energy is planned to be 250 GeV.}, is a promising facility for SUSY
searches. The capability of the ILC for determining exclusion/discovery limits
for the $\widetilde{\tau}$ in a model-independent way is shown in this paper, together
with an overview of the current state-of-the-art.
Results of the last studies of $\widetilde{\tau}$ pair-production at the ILC are presented,
showing the improvements with respect to previous results\footnote{Talk presented at the International Workshop on Future Linear Colliders (LCWS2021),  15-18 March 2021. C21-03-15.1.}.  
\end{abstract}

\begin{section}{Introduction}
  Supersymmetry (SUSY)~\cite{Martin:1997ns}\cite{Wess:1974tw}\cite{Nilles:1983ge}\cite{Haber:1984rc}\cite{Barbieri:1982eh}
    is one of the most promising candidates for new physics. It could explain or
  at least give some hint at solutions to current problems of the Standard Model (SM), such as the gauge hierarchy
  problem, the nature of Dark Matter or the possible theory-experiment discrepancy of the muon magnetic moment. 
  SUSY is a symmetry of
  spacetime relating fermions and bosons. For every SM particle it introduces a superpartner with the
  same quantum numbers except for the spin. The spin differs by half a unit from the value of its SM partner.
  A new parity, R-parity, is commonly introduced in SUSY, which has a crucial impact in SUSY phenomenology.
  R-parity takes an even value for SM particles and odd value for the SUSY ones.
  Multiplicative R-parity conservation~\footnote{The introduction and conservation of
  this symmetry is inspired by flavour physics constraints since the most general SUSY Lagrangian induces
  flavour-changing neutral interactions that are avoided imposing R-parity conservation.}, assumed in
  most of the SUSY models, implies that the SUSY particles are always created in
  pairs and that the lightest SUSY particle (the LSP) is stable and,  when cosmological constraints are taken into account,
  also neutral.
  An important point in this kind of studies is the fundamental SUSY principle stating that each
  SUSY particle couples as its corresponding SM particle. This allows to know the cross sections
  for SUSY pair production solely from the knowledge of initial centre-of-mass energy of the collider and the masses of the involved SUSY
  particles.
\end{section}

\begin{section}{SUSY searches}
  Considerable efforts have been and are being devoted to the search of SUSY at different facilities.
  Searches at hadron colliders, such as the LHC, are mainly sensitive to the production of coloured particles,
  gluino and squarks. They are most probably the heaviest ones. 
  The search of the lighter colour-neutral SUSY states, such as sleptons, charginos or neutralinos, at
  hadron colliders is challenged by the much smaller cross sections,
  and high backgrounds. Mass limits have been obtained at the LHC, but they
  are only valid if many constraints on the model parameters are fulfilled.
  Lepton colliders, like LEP, have a higher sensitivity to the production of colour-neutral SUSY states, but
  the searches up to now were limited by the beam energy. Limits computed at these facilities are however
  valid for any value of the model parameters not shown in the exclusion plots.
  The future International Linear Collider (ILC)~\cite{Behnke:2013xla}, an electron-positron collider operating at centre-of-mass
  energies of $250-500$\,GeV and with upgrade capability to $1$\,TeV, is seen as an ideal environment for SUSY studies.
  SUSY searches at the ILC would profit from the high electron and positron beam polarisations,
  80$\%$ and 30$\%$ respectively, a well defined initial state (in 4-momentum and spin configuration),
  a clean and reconstructable final state, near absence of pile-up, hermetic detectors (almost
  4$\pi$ coverage) and trigger-less operation, which is a huge advantage for precision measurements and
  unexpected signatures.
\end{section}  

\begin{section}{Motivation for $\widetilde{\tau}$ searches}
  For evaluating the power of SUSY searches at future facilities,
  it is beneficial to focus on the lightest particle in the SUSY spectrum that could be accessible. Since
  the cosmological constraints requires a neutral and colourles LSP, the next-to-lightest SUSY
  particle, the NLSP,  would be the first one to be detected. The NLSP can only decay to the LSP and
  the SM partner of the NLSP (or {\it via} it's SM partner, if the LSP-NLSP mass-difference is smaller than the mass
  of the SM partner). This already makes the NLSP production special: heavier states might well decay in
  cascades, and thus have signatures that depend strongly on the model.
  Furthermore, there is only a finite set of sparticles that could be the NLSP, so a systematic search
  for each possible case is feasible.
  This also means that one can a priori estimate which will be the most difficult case, namely the
  NLSP that combines small production cross-section with a difficult experimental signature.
  The $\widetilde{\tau}$ satisfies both these conditions. Therefore, studies  of $\widetilde{\tau}$ production
  might be seen as the way to determine the guaranteed discovery or exclusion reach for SUSY: any other NLSP
  would be easier to find.
  
  The $\widetilde{\tau}$ is the super-partner of the ${\tau}$. Like for any other fermion or sfermion, there
  are two weak hyper-charge states, $\widetilde{\tau}_{R}$ and $\widetilde{\tau}_{L}$. For the fermions the
  chiral symmetry assures that both weak hyper-charge states are degenerate in mass. However this symmetry
  does not apply to sfermions, since they are scalars, rather than fermions. Hence there is no reason to
  expect that  $\widetilde{\tau}_{R}$ and $\widetilde{\tau}_{L}$ would have the same mass.
  Furthermore, mixing between the weak hyper-charge states
  yields the physical states. The strength of the couplings involved in the mixing of states depend on the fermion mass and hence only
  the third generation of the sleptons, $\widetilde{\tau}$, will mix~\footnote{This is also the case for the
  squarks, where the third generation, the stop the and the sbottom, will also mix.}. As a consequence of the mixing the lightest
  $\widetilde{\tau}$, $\widetilde{\tau}_{1}$, would most likely  be the lightest slepton, due to the seesaw
  mechanism. The mass of the lightest physical (mixed) state would be smaller than the mass of
  any un-mixed weak hyper-charge state. The cross section of the $\widetilde{\tau}$ also differs from the
  one of the ${\tau}$, not only due to phase-space limitations - the $\widetilde{\tau}$ being more massive
  than the ${\tau}$ - but also due to the mixing.
  In $e^+e^{-}$ colliders, assuming R-parity conservation, the $\widetilde{\tau}$ will be pair-produced, with
  contribution of the s-channel only, via $Z^{0}$/$\gamma$ exchange. The strength of the  $Z^{0}$/$\gamma$ $\widetilde{\tau}$ $\widetilde{\tau}$ coupling depends on the $\widetilde{\tau}$ mixing, reaching its minimum value when
  the coupling $\widetilde{\tau}_{1}$ $\widetilde{\tau}_{1}$ $Z^{0}$ vanishes, which will lead to the worst possible
  scenario in $\widetilde{\tau}$ and, in general, in slepton searches. 
  Another property making the search of  $\widetilde{\tau}$ the worst case, is the fact that its SM partner
  is unstable. It decays before it can be detected, and, as a further complication, some of its decay products are
  undetectable neutrinos.
  This on one hand  makes the identification 
  more difficult than the direct decay to electrons or muons, and on the other hand,
  since the decay products are only partially detectable, that blurs kinematic signatures.
  The search of a light $\widetilde{\tau}$ is also theoretically motivated: SUSY models with a light
  $\widetilde{\tau}$ can accommodate the observed relic density, by enhancing the $\widetilde{\tau}$-neutralino coannihilation process. 
\end{section}

\begin{section}{Limits at LEP, LHC and previous ILC studies}
  The most model-independent limit on the $\widetilde{\tau}$ mass comes from the LEP experiments~\cite{LEPSUSYWG/04-01.1}.
  They set a minimum value that ranges from 87 to 93 GeV depending on the mass difference between the $\widetilde{\tau}$ and
  the neutralino, not smaller than 7 GeV. These limits, shown in figure~\ref{lep_limits}, are valid for any mixing and any value of the model-parameters, other than the two masses explicitly shown in the plot.

  \begin{figure}[htbp]
    \centering
    \includegraphics [width=.4\textwidth]{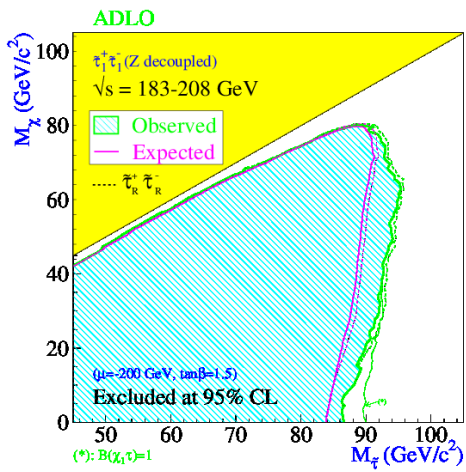}
    \caption{95$\%$ CL exclusion limits for $\widetilde{\tau}$ pair production obtained combining data collected at the four LEP experiments with
      energies ranging from 183 GeV to 208 GeV. From~\cite{LEPSUSYWG/04-01.1}.}
    \label{lep_limits}
  \end{figure}
  
  An analysis by the DELPHI experiment, targeted at low mass differences, excludes a $\widetilde{\tau}$ with mass below 26.3 GeV, for any mixing, and any mass difference larger than the $\tau$ mass~\cite{Abdallah:2003xe}.
  At the LHC, ATLAS and CMS have determined limits on the $\widetilde{\tau}$ mass, analysing data from Run 1 and
  Run 2~\cite{Aad:2019byo}~\cite{CMS:2019eln}.
  These limits, however, are only valid under certain assumptions. Both experiments assume $\widetilde{\tau}_{R}$
  and $\widetilde{\tau}_{L}$ to be mass-degenerate. This is a very unlikely scenario, the running of the $\widetilde{\tau}_{R}$
  and $\widetilde{\tau}_{L}$ masses from the GUT scale to the weak scale follows renormalisation group equations with $\beta$-functions
  that are inevitably different for the two weak hyper-charge states.
  They also assume that there is no mixing between the weak hyper-charge eigenstates, which is again very improbable.
  The mixing will yield to cross section of the lightest physical state smaller than that of any unmixed state.
  Putting together $\widetilde{\tau}_{R}$ and $\widetilde{\tau}_{L}$ by adding the cross sections, ATLAS excludes
  $\widetilde{\tau}$ masses between approximately 120 and 390~\,GeV for a nearly massless neutralino\footnote{100$\%$
    decay to $\widetilde{\tau}$ and neutralino is assumed, as it is in the analysis presented in this paper}.
  Under the same conditions, CMS extends the lower limit to 90~\,GeV closing the gap with the LEP limit. Analysis of a pure
  $\widetilde{\tau}_{L}$ pair production set limits between 150 and 310~\,GeV from ATLAS data and up to 125~\,GeV from CMS; both limits  again
  assume a nearly massless neutralino.
  The future HL-LHC should provide an improvement on the $\widetilde{\tau}$ limits provided by ATLAS and
  CMS, not only because of an increase of the luminosity but also because of an expected gain in sensitivity
  to direct $\widetilde{\tau}$ production due to the use of different analysis methods.
  Simulation studies have already been performed in both
  experiments~\cite{ATLAS:2018diz}~\cite{CMS:2018imu}. Upper limits for $\widetilde{\tau}$ masses
  are indeed increased by about 300~\,GeV, but they suffer from the same constraints as the previous
  studies. ATLAS adds limits for pure $\widetilde{\tau}_{R}$ pair production, that could be considered
  the closest case to the physical lightest $\widetilde{\tau}$ since it is likely to be the lightest of the two
  weak hyper-charge states and is the one with the lowest cross section. These limits, presented in figure~\ref{atlas_hllhc},
  show that no discovery
  potential is expected in this case, only exclusion potential. They do not have exclusion potential for $\widetilde{\tau}$ co-annihilation
  scenarios, a highly motivated scenario if SUSY is to provide a viable DM candidate:
  Such a scenario requires that the   $\widetilde{\tau}$-LSP mass difference is small, $\lesssim$ 10 GeV. 
  $\widetilde{\tau}$ searches at the ILC have been also performed in previous studies~\cite{Berggren:2013vna}.
  They assume an integrated luminosity of 500~\,fb$^{-1}$ at $\sqrt{s}=500$\,GeV and average
  beam polarisations of $P(e^{-},e^{+})=(+80\%,-30\%)$. The same beam polarisations are used in the
  current studies, since the signal to background ratio is favoured, but the luminosity is increased
  to the one corresponding the the foreseen running scenario, 1.6~\,ab$^{-1}$.
  The limits presented in that study do not have a dedicated
  analysis for low mass differences between the $\widetilde{\tau}$ and the LSP, $\Delta M$, and are only
  valid down to $\Delta M$ 3-4~\,GeV. The exclusion limit goes up to 240\,GeV with a discovery potential
  up to 230\,GeV for large mass differences.
  
  \begin{figure}[htbp]
    \hspace*{-3cm}
    \includegraphics [width=1.0\textwidth]{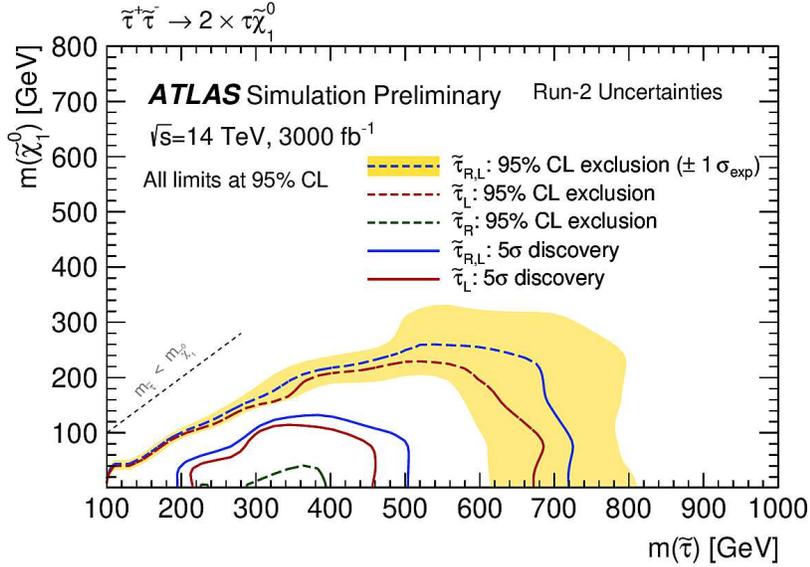}
    \vspace*{-3cm}
    \caption{95$\%$ CL exclusion and discovery potential for $\widetilde{\tau}$ pair production at the HL-LHC, assuming
    $\widetilde{\tau}_{L}^{+}\widetilde{\tau}_{L}^{-}$ + $\widetilde{\tau}_{R}^{+}\widetilde{\tau}_{R}^{-}$ production, $\widetilde{\tau}_{L}^{+}\widetilde{\tau}_{L}^{-}$ production or $\widetilde{\tau}_{R}^{+}\widetilde{\tau}_{R}^{-}$ production. From~\cite{ATLAS:2018diz}.}
    \label{atlas_hllhc}
  \end{figure}
  
\end{section}

\begin{section}{Conditions and tools}
  The study was done assuming an integrated luminosity of 1.6~\,ab$^{-1}$ at $\sqrt{s}=500$\,GeV
  with beam polarisations  $P(e^{-},e^{+})=(+80\%,-30\%)$, according to the H-20
  running scenario in the ILC500 benchmark~\cite{Barklow:2015tja} \footnote{$\sqrt{s}$=500\,GeV, total integrated
  luminosity 4\,ab$^{-1}$ with 1.6\,ab$^{-1}$ for $P(e^{-},e^{+})=(-80\%,+30\%)$ and 
  $P(e^{-},e^{+})=(+80\%,-30\%)$, 0.4\,ab$^{-1}$ for $P(e^{-},e^{+})=(+80\%,+30\%)$ and
  $P(e^{-},e^{+})=(-80\%,-30\%)$}. The polarisation was selected due to the
  increase of the signal to background ratio, as will be shown in the description of the
  analysis.
  The study assumes R-parity conservation and a 100$\%$ decay of the $\widetilde{\tau}$ to ${\tau}$
  and the lightest neutralino, the LSP in this case.
    In order to select the worst scenario, the $\widetilde{\tau}$ mixing angle was set to 53 degrees,
  corresponding to the lowest cross section  due to the suppression of the s-channel with Z exchange
  in the $\widetilde{\tau}$ pair production.
  The SGV fast detector simulation~\cite{Berggren:2012ar}, adapted to the ILD  concept~\cite{ILD:2020qve} at ILC, was used for detector simulation and event
  reconstruction. Signal events were generated inside SGV using Pythia 6.422~\cite{Sjostrand:2006za}.
  The generated background event samples were those of the standard ``DBD'' production~\cite{Behnke:2013lya}. They were generated
  with Whizard 1.95~\cite{Kilian:2007gr}, and were written in stdhep format. These files were read by SGV, and passed through the same detector simulation and reconstruction as the signal samples.
  The relevant information of the reconstructed events were written to Root files.
\end{section}

\begin{section}{Signal characterisation}
  Assuming R-parity conservation and assuming that the $\widetilde{\tau}$ is the NLSP, $\widetilde{\tau}$'s
  will be produced in pairs via $Z^{0}$/$\gamma$ exchange in the s-channel and they will decay to
  a ${\tau}$ and an LSP (assuming mass differences above the mass of the ${\tau}$, as is done
    in this study).
  The LSP, as already mentioned, is stable and weakly interacting, hence it will leave the detector
  without being detected. The ${\tau}$, with a lifetime of the order of 2.9 x 10$^{-13}$ s, will decay
  before leaving any signal in the detectors. The only detectable activity in the signal events is therefore
  the decay products of the two ${\tau}$'s.
  Signal events are then characterised by a large missing energy and momentum, not only due to the
  invisible LSPs but also to the neutrinos from both ${\tau}$-decays. Since the $\widetilde{\tau}$'s are scalars
  and hence isotropically produced, these events have a large fraction of the detected activity in the
  central region of the detector. The $\widetilde{\tau}$'s must also be rather heavy, so they will not have a large boost in the lab-frame,
  and since the LSP is also quite heavy, the direction of the $\widetilde{\tau}$ does not  strongly correlate to that of
  the visible $\tau$ after the  decay. As a consequence the two $\tau$-leptons are expected to go in directions quite independent
  of each other resulting in events with un-balanced transverse momentum, large angles between the two $\tau$-lepton directions and absence of
  forward-backward asymmetry.
  These properties are  however not necessarily present in any event - the two $\tau$'s could accidentally happen
  to be back to back, for example. 
\end{section}

\begin{section}{Main background sources}
  The main sources of background, given the generic signal topology, i.e. two $\tau$'s and an anseen
  recoil system, are SM processes with real or fake missing energy. They can be classified into
  ``irreducible'' and ``almost irreducible'' sources. The first are events with two $\tau$'s
  and real missing energy, i.e. neutrinos. The main contribution to this group are ZZ events
  with one Z decaying to two neutrinos and the other to two $\tau$'s, and fully leptonic WW events,
  where at most one of the W's decays to $\tau$ and neutrino.
  ZWW and ZZZ events decaying to two $\tau$'s and four neutrinos
  are not an issue due to their low cross sections.
  The second group of events are those which are not really two $\tau$'s and neutrinos, but after reconstruction looks very similar.
  They are events with two soft $\tau$-jets, with two other leptons plus true
  missing energy or two $\tau$'s plus fake missing energy.
  The main sources for events with true missing energy in this group are $\tau$
  pair production, with the $\tau$'s decaying such that most energy goes to the neutrinos, ZZ events where one of the Z's decays to an electron or a muon
  pair and the other one to neutrinos, and WW events
  with each W decaying to an electron or muon and a neutrino.
  The background with fake missing energy comes mainly from $\tau$ pair
  production with Initial State Radiation (ISR) at very low angles, events with two $\tau$'s and
  two very low angle electrons (below the acceptance of the BeamCAL) in the final state and events where two $\tau$'s are produced
  by a $\gamma\gamma$ interaction and not from an $e^+e^-$ one; in that case there
  is not really missing energy but an initial state with much less energy
  than that of the electron-positron interactions.
\end{section}

\begin{section}{General cuts}
  
  Taking into account the signal signature and the main background sources,
  different cuts have been designed in order to separate the signal from
  the background.

  Since the study was focused on small differences between the $\widetilde{\tau}$ and LSP
  masses\footnote{Larger mass differences were also analysed in order
  to cross-check and try to improve the limits from the previous studies.},
  $M_{\tau}$ $<$ $\Delta M$ $<$ 11\,GeV, the absence of signal in the calorimeter closest to the beam pipe
  (the BeamCAL) was required as a pre-selection step before applying the following cuts. 
  
  The first group of cuts are those in properties that the
  $\widetilde{\tau}$-events {\it must} have. Since the two LSP's from the
  $\widetilde{\tau}$-decays are invisible to the detector, signal events
  have to have a missing energy greater than two times the mass of the
  LSP and the visible mass can not be bigger than this quantity. Also a
  cut in the maximum total momentum, smaller than 70$\%$ the beam momentum
  is applied for the same reason. The multiplicity of the event can also
  be constrained taking into account that the visible part comes only from
  the decays of the two $\tau$'s and maybe an ISR photon. For that reason
  the number of charged particles is asked to be between 2 and 6, with
  only 2 or 3 clusters identified as $\tau$'s and a total charge between
  -1 and 1. An specific algorithm for $\tau$-identification was also applied.
  This algorithm consists in a first set of conditions requiring to have a pattern
  of charged tracks typical for $\tau$-decay, {\it viz.} exactly two jets
  (obtained with the DELPHI tau-finder~\cite{Abdallah:2003xe}) with
  charged particles, 1 or 3 charged particles in each charged jet, 
  jet-charge $\pm$1, and opposite charge between both jets. A set of conditions
  is devoted to the reduction of background from sources with
  leptons not from $\tau$-decays. To reduce the background of single W
  production in e$\gamma$ events, with W decaying to $\tau$ and neutrino,
  none of the jets should consist of a single positron (this cut takes
  into account the polarisation selected for the study). This background
  together with the background from $WW \rightarrow e\nu_{e} \mu\nu_{\mu}$ and
  from $\gamma\gamma$ events with a beam-remnant deflected to larger
  angles is further reduced by rejecting those events in which the most energetic
  jet consists of a single electron. The two charged jets were also required to
  neither be made by single leptons with the same flavour nor to have one hadronic
  jet and one leptonic.
  This algorithm reduces the signal efficiency by 38$\%$ but with
  a reduction of the background of the order of 94$\%$, depending on the
  region of the SUSY parameter space one is working with.
  The last cut in this first group of cuts  is on the maximum momentum of the jets. Since
  the $\widetilde{\tau}$-decay is a two body decay, it is possible to
  determine the maximum and minimum momentum of each of the  decay products as
  a function of the $\widetilde{\tau}$ mass, the mass of LSP
  and the centre-of-mass energy of the collider. The cut in the minimum momentum
  can not be applied due to the presence of neutrinos in the $\tau$ decay, with
  the corresponding decrease of observable momentum. The maximum value can be used even if
  it is smeared by the missed neutrinos. The expression for the maximum
  jet momentum is given by:

  \begin{equation}
    P_{max} = \frac{\sqrt{s}}{4}(1-\left ( \frac{M_{LSP}}{M_{\widetilde{\tau}}} \right )^2)(1+\sqrt{(1-\frac{4M_{\widetilde{\tau}}~^2}{s}})
  \end{equation}

  Excluding the cut applied by the $\tau$-identification algorithm, the signal
  efficiency for each of the cuts is at least 95$\%$ at all model points.

  A second group of cuts is based on those properties that the
  $\widetilde{\tau}$-events {\it might} have, but will {\it rarely} be
  present in background events. As already pointed out, the  $\widetilde{\tau}$'s are
  scalars, and therefore isotropically produced, while the backgrounds are
  either fermions or vector bosons, and tend to be produced at small angles to the beam
  axis. This allows to set cuts requiring events with high missing transverse
  momentum, large acoplanarity, high angles to the beam and high
  values of the variable $\rho$.
  The latter is calculated by first projecting the event on the x-y
  (transverse) plane, and calculating the thrust axis in that plane. $\rho$
  is then the transverse momentum (in the plane) with respect to the
  thrust axis.
  The cut in $\rho$ helps to reject events with two $\tau$'s back-to-back in the transverse
  projection with the {\it visible} part of the decay of one of the $\tau$'s in the direction of its parent,
  while the other $\tau$ decays with the {\it invisible} $\nu$ closely aligned with the direction of its parent.
  These events fake the signal topology,
  having a large missing transverse momentum and high acoplanarity, but would have
  a small value of $\rho$. The values at which the cut in
  these properties is set depends on the $\widetilde{\tau}$ mass and the mass difference
  between the $\widetilde{\tau}$ and the LSP. Cutting in these properties has a certain
  cost in efficiency but improves the signal-to-background ratio. 

  The third group of cuts uses properties of some of the ``almost irreducible'' sources of
  background. WW events with each of the W's decaying to a lepton (not $\tau$) and a neutrino are
  highly forward-backward asymmetric; they can be almost entirely removed by requiring the sum of the product of the
  charge and the cosine of the polar angle of the two most energetic jets to be above -1.
  ZZ events with one Z decaying to two neutrinos and the second one to a electron or muon
  pair are highly suppressed demanding a visible mass more than 4~\,GeV from the Z mass,
  since the visible mass in those events equals the Z mass quite precisely.

  A last cut is based on a property that the signal often {\it does not} have, {\it viz.} sizeable energy
  detected at low angles to the beam. Events with more than 2 GeV detected at angles lower than 20
  degrees to the beam axis are therefore rejected.
  This cut is however not useful for small mass differences.

  After applying these cuts the main sources of remaining background are WW events with each
  W decaying to $\tau\nu$ and events with four fermions in the final state coming from
  $\gamma\gamma$ interactions, mostly $\tau\tau$ events.

  The selected polarisation plays an important role in the capability of excluding/discovering
  the different regions of the SUSY space.
  Table \ref{tab:polarisations} shows the number of signal and background events for a specific spectrum point for
  the two main ILC running polarisations and for unpolarised beams.
  Since the polarisation of the $\tau$ coming from the $\widetilde{\tau}$ decays was not considered in
  this study, the difference in the number of signal events comes only from the dependence of the
  cross section on the polarisation. This is also the main factor for the difference in WW events, ee $\rightarrow\tau\nu\tau\nu$.
  One can see that the signal-to-background ratio is clearly enhanced in the selected polarisation.
  Taking the definition of exclusion at 95$\%$ CL as $S > 2\sqrt{S+B}$, with S and B the number of signal and background
  events respectively, it is also shown that unpolarised beams would allow neither exclusion nor discovery. 
  Polarisation is not only important in the enhancement of the signal over background but also
  plays an important role in the parameter determination.
  \begin{table}[htbp]
    \centering
    \caption{Remaining signal and background events after the application of the selection cuts for
      M$_{\widetilde{\tau}}$=47~\,GeV and mass difference with the LSP of 10~\,GeV. 
    }
    \label{tab:polarisations}
    \begin{tabular}{lcccc}
      \hline\hline
      \addlinespace[2pt]
      Polarisation & Signal & ee $\rightarrow\tau\nu\tau\nu$ & $\gamma\gamma \rightarrow \tau\tau$ll & $\gamma\gamma \rightarrow$ llll\\[2pt]
      \hline
      \addlinespace[2pt]
      $P(e^{-},e^{+})=(+80\%,-30\%)$ & 7.4 & 1.7 & 0.2 & 0.02 \\
      $P(e^{-},e^{+})=(-80\%,+30\%)$ & 5.7 & 28  & 0.2 & 0.02 \\
      Unpolarised                     & 6   & 12  & 0.2 & 0.02  \\
      \hline
      \addlinespace[2pt]
    \end{tabular}
  \end{table}
   
\end{section}

  The cuts described above are mainly suited for mass differences up to 3~\,GeV.
  When the mass difference is between 3~\,GeV and the mass of the $\tau$ \footnote{For mass differences
    below the mass of the $\tau$ the lifetime of the $\widetilde{\tau}$ increases exponentially and
    the study has to be done based on a signature of long-lived particles travelling through the
    detector.} the kinematic of the signal events is very close to that of the $\gamma\gamma$ background
  events and the described cuts are not enough for discovering/excluding the signal.
  An additional cut was done based on the Initial State Radiation photons (ISR). Events with isolated photons
  with high transverse momentum were selected, allowing to extension of the limits into the region under study.
  This cut is effective against the remaining $\gamma\gamma$ background because these events become candidates due
  to fake missing transverse momentum. If the presence of an ISR is requested, the incoming electron or positron that
  emitted the ISR must have recoiled against the ISR.
  Since this is a
  scattering process, not an annihilation one, the electron (positron) is still present in the final state.
  Therefore, if it is required to see a high transverse momentum ISR, the final state
  electron (positron) will have acquired a transverse momentum big enough to be deflected into the BeamCAL, and
  thus to have been rejected already at the pre-selection stage. On the other hand, if the ISR was emitted from
  an electron or positron that was subsequently annihilated into a Z, as is the case for the signal process,
  the transverse momentum of the ISR is included
  in the decay products of the Z, and no signal is expected in the BeamCAL.

\begin{section}{Exclusion/discovery limits}
  The exclusion and discovery limits extracted from this study are shown in figure~\ref{exclusion_mstauvsmneu_prev}.

  \begin{figure}[htbp]
    \centering
    \includegraphics [width=0.7\textwidth]{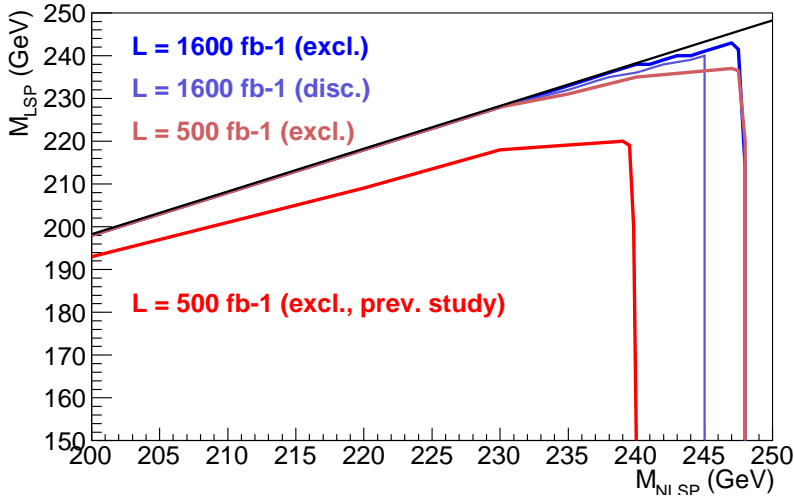}
    \caption{Exclusion and discovery $\widetilde{\tau}$ limits as a function of the $\widetilde{\tau}$ and LSP masses. Exclusion limits from previous ILC studies are also shown, as well as an extrapolation of the currents limits to 500~\,fb$^{-1}$ total integrated luminosity.}
    \label{exclusion_mstauvsmneu_prev}
  \end{figure}
  
  They assume the lightest $\widetilde{\tau}$ to be the NLSP and the lightest neutralino the LSP,
  and are valid for any  $\widetilde{\tau}$ mixing angle. Results from previous ILC studies,
  computed for 500~\,fb$^{-1}$ total integrated luminosity, are also shown for comparison, as well
  as an extrapolation of the current results from 1.6~\,ab$^{-1}$ to 500~\,fb$^{-1}$. The
  comparison of these two curves shows that the extension of the limits is not only due to an increase of the
  total integrated luminosity but also to an improvement of the analysis. The main reason of this improvement
  is the application of individual limits depending on the $\widetilde{\tau}$ mass and the mass difference.
  The previous studies were only making a difference for mass differences above or below 10~\,GeV and
  were not optimised for the low mass difference region. Another difference in the analysis is
  a change in the ${\tau}$-identification algorithm, excluding events with two jets consisting of single leptons of
  the same flavour or one jet being hadronic and other leptonic, which was found to be  necessary for the exclusion/discovery
  of some points.
  It is also relevant to compare these results with the current $\widetilde{\tau}$ limits coming from
  LEP. Figure~\ref{exclusion_mstauvsmneu} shows this comparison. LEP limits are also valid for any value of the not shown
  model parameters.

  \begin{figure}[htbp]
    \centering
    \includegraphics [width=0.5\textwidth]{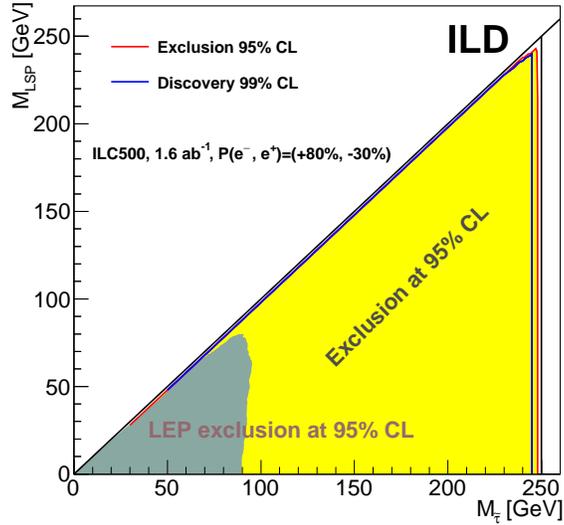}
    \caption{Exclusion and discovery $\widetilde{\tau}$ limits from the current studies compared to the ones from LEP studies.}
    \label{exclusion_mstauvsmneu}
  \end{figure}
  
  The projection of the limits in the M$_{\widetilde{\tau}}$-$\Delta$M plane is shown in figure~\ref{exclusion_mstauvsdm}. The
  region for mass differences below the mass of the $\tau$, not included in the current study, is shown for
  completeness. In the region with $\Delta$M larger than M$_{\tau}$ exclusion and discovery ILC limits
  are compared to the ones from LEP. Since the LHC limits are highly model-dependent, the comparison in
  this case have to be taken with care. Limits considering only the $\widetilde{\tau}_{R}$-pair production
  are shown, as, while still being optimistic, they are the closest to the ones expected for the lightest
  $\widetilde{\tau}$ at minimal cross-section.
  For the region with $\Delta$M smaller than M$_{\tau}$ results from LEP and LHC are shown.
  LEP studies cover not only the region where the $\widetilde{\tau}_{1}$ travels through the
  detector without decaying but also the region with decays at a certain distance from the production
  vertex. In those regions acoplanar leptons, tracks with large impact parameters and kinked tracks
  are looked for, depending on the $\widetilde{\tau}_{1}$ lifetime~\cite{LEPSUSYWG/02-05.1}~\cite{LEPSUSYWG/02-09.2}.
  Figure~\ref{exclusion_mstauvsdm_extrap} extends the previous one adding the extrapolation of the ILC limits for the scenarios
  with centre-of-mass energy 250~\,GeV and 1~\,TeV.

  \begin{figure}[htbp]
    \centering
    \includegraphics [width=0.5\textwidth]{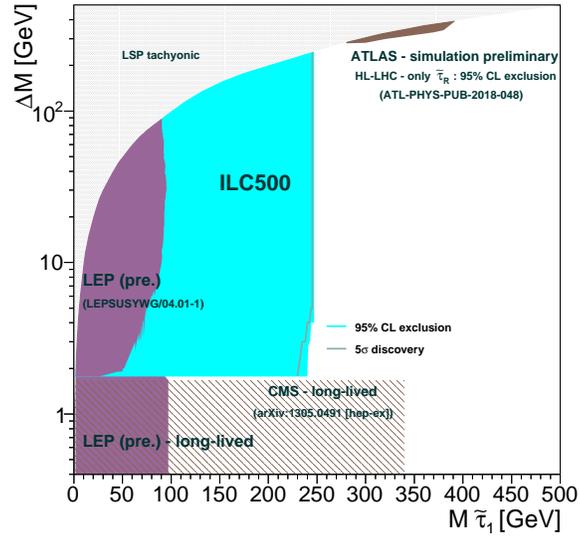}
    \caption{$\widetilde{\tau}$ limits in the M$_{\widetilde{\tau}}$-$\Delta$M plane. ILC results from the current studies are shown together with limits from LEP and LHC. The region with mass differences below the mass of the ${\tau}$ is also shown with LEP and LHC results, even if it is not covered by this study.}
    \label{exclusion_mstauvsdm}
  \end{figure}
  
  \begin{figure}[htbp]
    \centering
    \includegraphics [width=0.5\textwidth]{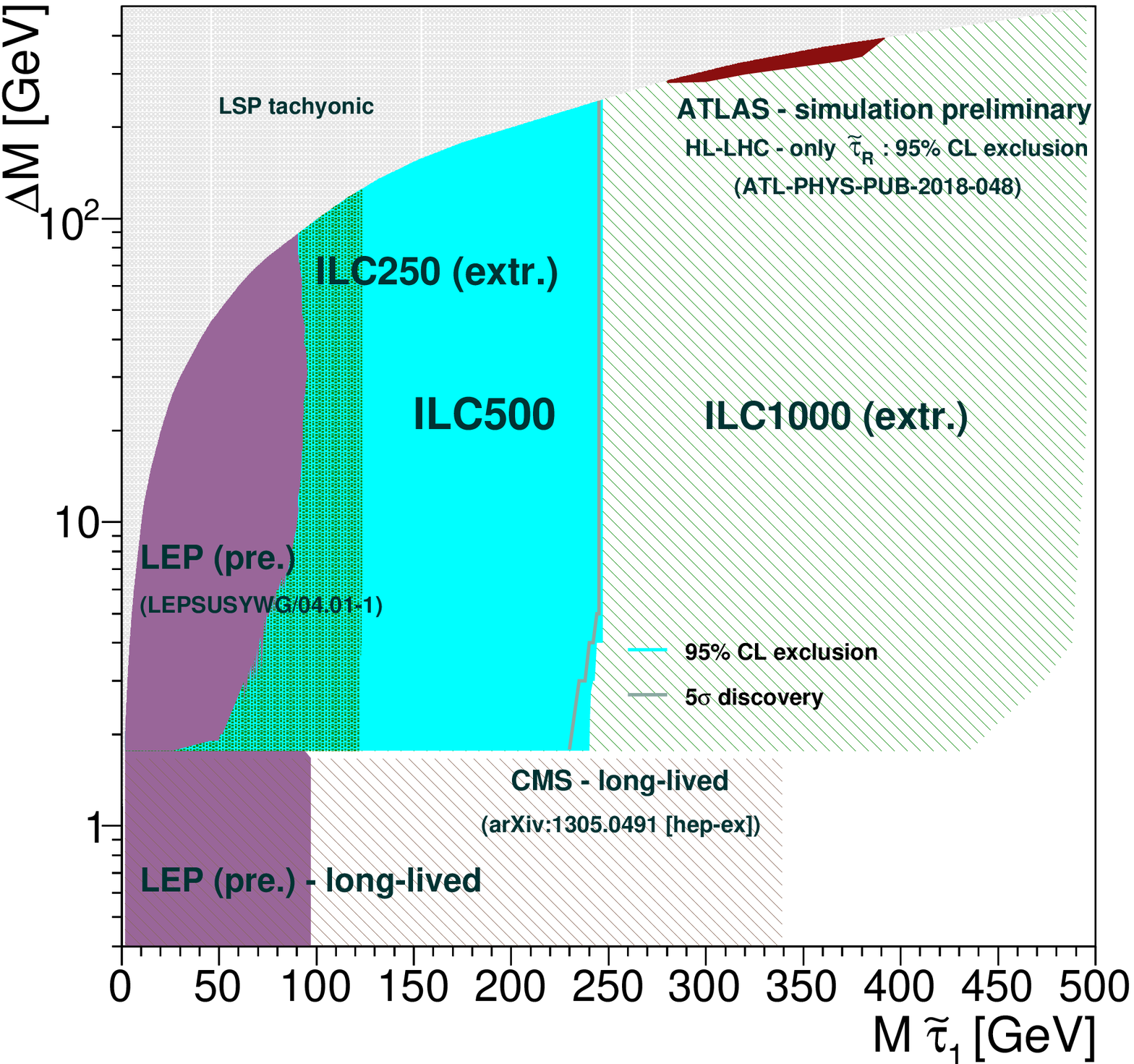}
    \caption{$\widetilde{\tau}$ limits in the M$_{\widetilde{\tau}}$-$\Delta$M plane with the extrapolation of
      the ILC current results to the ILC 250~\,GeV and 1~\,TeV running scenarios.}
    \label{exclusion_mstauvsdm_extrap}
  \end{figure}

\end{section}

\begin{section}{Outlook and conclusions}
The capability of the ILC for excluding/discovering $\widetilde{\tau}$-pair production up to a
few GeV below the kinematic limit, without model dependencies and even in the worst scenario,
has been shown.
The study has been done assuming the $\widetilde{\tau}$ mixing angle to be the one corresponding to
the lowest cross section for unpolarised beams. This is also the mixing angle that
gives the smallest number of signal events when simply combining the samples with 
polarisations $P(e^{-},e^{+})=(+80\%,-30\%)$ and $P(e^{-},e^{+})=(-80\%,-30\%)$ with equal
integrated luminosity, as it is planned in the ILC running scenarios. However, due to the
clear enhancement of the signal-to-background ratio with the polarisation $P(e^{-},e^{+})=(+80\%,-30\%)$, 
as shown in table \ref{tab:polarisations}, only this dataset was used for the calculation of
the limits. The study will be extended taking into account the contribution of both polarisations. In this extension we will
consider different $\widetilde{\tau}$ mixing angles
for confirming the one corresponding to the worst scenario.
Without considering the polarisation of the ${\tau}$ coming from the $\widetilde{\tau}$ decay, as it
is done in the present study, the number of detected signal events
for each mixing angle and each beam polarisation depends only on the cross section for
$\widetilde{\tau}$-pair production in those conditions. However the signal efficiency is
affected by the ${\tau}$ polarisation due to the effect on the momentum distribution of the ${\tau}$-decays,
being softer or harder depending on the neutralino mixing angle. This effect will be also
considered in the extension of the study, being an important point in the determination of the
worst scenario.
The calculation of the exclusion/discovery limits in the region with mass differences below the
${\tau}$ mass, meaning an exponential increase of the $\widetilde{\tau}$ lifetime and consequently
a study of long-lived particles going through or decaying in different parts of the detector, is
also foreseen.

\end{section}
   
\begin{section}{Acknowledgements}
  We would like to thank the LCC generator working group for producing the Monte Carlo samples used in this study.
  We also thankfully acknowledge the support by the the Deutsche Forschungsgemeinschaft (DFG, German Research Foundation) under Germany’s Excellence Strategy EXC 2121 ``Quantum Universe'' 390833306.
  This work has benefited from computing services provided by the German National Analysis Facility (NAF)~\cite{Haupt_2010}.
\end{section}


\medskip

\begin{thebibliography}{1}
  
\bibitem{Martin:1997ns}
S.~P.~Martin,
%
``A Supersymmetry primer,''
Adv. Ser. Direct. High Energy Phys. \textbf{18} (1998), 1-98
[arXiv:hep-ph/9709356 [hep-ph]].

\bibitem{Wess:1974tw}
J.~Wess and B.~Zumino,
%
``Supergauge Transformations in Four-Dimensions,''
Nucl. Phys. B \textbf{70} (1974), 39-50

\bibitem{Nilles:1983ge}
H.~P.~Nilles,
%
``Supersymmetry, Supergravity and Particle Physics,''
Phys. Rept. \textbf{110} (1984), 1-162

\bibitem{Haber:1984rc}
H.~E.~Haber and G.~L.~Kane,
%
``The Search for Supersymmetry: Probing Physics Beyond the Standard Model,''
Phys. Rept. \textbf{117} (1985), 75-263

\bibitem{Barbieri:1982eh}
R.~Barbieri, S.~Ferrara and C.~A.~Savoy,
%
``Gauge Models with Spontaneously Broken Local Supersymmetry,''
Phys. Lett. B \textbf{119} (1982), 343

\bibitem{Behnke:2013xla}
T.~Behnke, J.~E.~Brau, B.~Foster, J.~Fuster, M.~Harrison, J.~M.~Paterson, M.~Peskin, M.~Stanitzki, N.~Walker and H.~Yamamoto,
[arXiv:1306.6327 [physics.acc-ph]].

  
\bibitem{LEPSUSYWG/04-01.1}
  LEP SUSY Working Group, ALEPH, DELPHI, L3 and OPAL Collaborations,
  ``Combined LEP Selectron/Smuon/Stau Results, $183$-$208$ GeV,''
  LEPSUSYWG/04-01.1,
  http://lepsusy.web.cern.ch/lepsusy/www/sleptons\_summer04/slep\_final.html
  
  
\bibitem{Abdallah:2003xe}
J.~Abdallah \textit{et al.} [DELPHI],
%
``Searches for supersymmetric particles in e+ e- collisions up to 208-GeV and interpretation of the results within the MSSM,''
Eur. Phys. J. C \textbf{31} (2003), 421-479
[arXiv:hep-ex/0311019 [hep-ex]].

\bibitem{Aad:2019byo}
G.~Aad \textit{et al.} [ATLAS],
%
``Search for direct stau production in events with two hadronic $\tau$-leptons in $\sqrt{s} = 13$ TeV $pp$ collisions with the ATLAS detector,''
Phys. Rev. D \textbf{101} (2020) no.3, 032009
[arXiv:1911.06660 [hep-ex]].

  

\bibitem{CMS:2019eln}
A.~M.~Sirunyan \textit{et al.} [CMS],
%
``Search for direct pair production of supersymmetric partners to the $\tau$ lepton in proton-proton collisions at $\sqrt{s}=$ 13 TeV,''
Eur. Phys. J. C \textbf{80} (2020) no.3, 189
[arXiv:1907.13179 [hep-ex]].


\bibitem{ATLAS:2018diz}
 [ATLAS],
 %
 ``Prospects for searches for staus, charginos and neutralinos at the high luminosity LHC with the ATLAS Detector,''
ATL-PHYS-PUB-2018-048.

\bibitem{CMS:2018imu}
 [CMS],
 %
 ``Search for supersymmetry with direct stau production at the HL-LHC with the CMS Phase-2 detector,''
CMS-PAS-FTR-18-010.

\bibitem{Berggren:2013vna}
M.~Berggren,
%
``Simplified SUSY at the ILC,''
[arXiv:1308.1461 [hep-ph]].
   
\bibitem{Barklow:2015tja}
T.~Barklow, J.~Brau, K.~Fujii, J.~Gao, J.~List, N.~Walker and K.~Yokoya,
%
``ILC Operating Scenarios,''
[arXiv:1506.07830 [hep-ex]].

\bibitem{Berggren:2012ar}
M.~Berggren,
%
``SGV 3.0 - a fast detector simulation,''
[arXiv:1203.0217 [physics.ins-det]].

\bibitem{ILD:2020qve}
H.~Abramowicz \textit{et al.} [ILD Concept Group],
%
``International Large Detector: Interim Design Report,''
[arXiv:2003.01116 [physics.ins-det]].

\bibitem{Sjostrand:2006za}
T.~Sjostrand, S.~Mrenna and P.~Z.~Skands,
%
``PYTHIA 6.4 Physics and Manual,''
JHEP \textbf{05} (2006), 026
[arXiv:hep-ph/0603175 [hep-ph]].

\bibitem{Behnke:2013lya}
T.~Behnke, J.~E.~Brau, P.~N.~Burrows, J.~Fuster, M.~Peskin, M.~Stanitzki, Y.~Sugimoto, S.~Yamada, H.~Yamamoto, H.~Abramowicz, \textit{et al.}
%
``The International Linear Collider Technical Design Report - Volume 4: Detectors,''
[arXiv:1306.6329 [physics.ins-det]].
  
\bibitem{Kilian:2007gr}
W.~Kilian, T.~Ohl and J.~Reuter,
%
``WHIZARD: Simulating Multi-Particle Processes at LHC and ILC,''
Eur. Phys. J. C \textbf{71} (2011), 1742
[arXiv:0708.4233 [hep-ph]].

\bibitem{LEPSUSYWG/02-05.1}
  LEP SUSY Working Group, ALEPH, DELPHI, L3 and OPAL Collaborations,
  ``Stable Heavy Charged Particles''
  LEPSUSYWG/02-05.1,
  http://lepsusy.web.cern.ch/lepsusy/www/stable\_summer02/stable\_208.html
  
\bibitem{LEPSUSYWG/02-09.2}
  LEP SUSY Working Group, ALEPH, DELPHI, L3 and OPAL Collaborations,
  ``Combined LEP GMSB Stau//Smuon/Selectron Results, 189-208 GeV''
  LEPSUSYWG/02-09.2,
  http://lepsusy.web.cern.ch/lepsusy/www/gmsb\_summer02/lepgmsb.html
  
\bibitem{Haupt_2010}
  A.~Haupt and Y.~Kemp
  ``The NAF: National analysis facility at DESY,''
  J. Phys.: Conf. Ser. \textbf{219} (2010), 052007

\end{thebibliography}

\end{document}